\documentclass[aps,prl,reprint,superscriptaddress,showpacs]{revtex4-1} 
\usepackage{graphicx}  
\usepackage{dcolumn}   
\usepackage{amssymb}
\usepackage{amsmath}
\usepackage{physics}
\usepackage{natbib}
\usepackage{color}
\usepackage{siunitx}

\begin{document}

\title{Closing of the Induced Gap in a Hybrid Superconductor-Semiconductor Nanowire}

\author{D.~Puglia}
\affiliation{Microsoft Quantum Labs Copenhagen and Center for Quantum Devices, Niels Bohr Institute, University of Copenhagen, Universitetsparken 5, 2100 Copenhagen, Denmark}
\affiliation{Institute of Science and Technology Austria, Am Campus 1, 3400 Klosterneuburg, Austria}
\author{E.~A.~Martinez}
\affiliation{Microsoft Quantum Labs Copenhagen and Center for Quantum Devices, Niels Bohr Institute, University of Copenhagen, Universitetsparken 5, 2100 Copenhagen, Denmark}
\author{G.~C.~M\'enard}
\affiliation{Microsoft Quantum Labs Copenhagen and Center for Quantum Devices, Niels Bohr Institute, University of Copenhagen, Universitetsparken 5, 2100 Copenhagen, Denmark}
\author{A.~P{\"o}schl}
\affiliation{Microsoft Quantum Labs Copenhagen and Center for Quantum Devices, Niels Bohr Institute, University of Copenhagen, Universitetsparken 5, 2100 Copenhagen, Denmark}

\author{S.~Gronin}
\affiliation{Microsoft Quantum Purdue, and Birck Nanotechnology Center, Purdue University, West Lafayette, IN, USA}

\author{G.~C.~Gardner}
\affiliation{Microsoft Quantum Purdue, and Birck Nanotechnology Center, Purdue University, West Lafayette, IN, USA}

\author{R.~Kallaher}
\affiliation{Microsoft Quantum Purdue, and Birck Nanotechnology Center, Purdue University, West Lafayette, IN, USA}

\author{M.~J.~Manfra}
\affiliation{Microsoft Quantum Purdue, and Birck Nanotechnology Center, Purdue University, West Lafayette, IN, USA}
\affiliation{Department of Physics and Astronomy, Purdue University, West Lafayette, IN, USA}
\affiliation{School of Materials Engineering, Purdue University, West Lafayette, IN, USA}
\affiliation{School of Electrical and Computer Engineering, Purdue University, West Lafayette, IN, USA}

\author{C.~M.~Marcus}
\affiliation{Microsoft Quantum Labs Copenhagen and Center for Quantum Devices, Niels Bohr Institute, University of Copenhagen, Universitetsparken 5, 2100 Copenhagen, Denmark}

\author{A.~P.~Higginbotham}
\email{Equal contribution, andrew.higginbotham@ist.ac.at}
\affiliation{Microsoft Quantum Labs Copenhagen and Center for Quantum Devices, Niels Bohr Institute, University of Copenhagen, Universitetsparken 5, 2100 Copenhagen, Denmark}
\affiliation{Institute of Science and Technology Austria, Am Campus 1, 3400 Klosterneuburg, Austria}
\author{L.~Casparis}
\email{Equal contribution, lucas.casparis@microsoft.com}
\affiliation{Microsoft Quantum Labs Copenhagen and Center for Quantum Devices, Niels Bohr Institute, University of Copenhagen, Universitetsparken 5, 2100 Copenhagen, Denmark}

\maketitle
\def\thefootnote{*}\footnotetext{These authors contributed equally.}

\textbf{
Hybrid superconductor-semiconductor nanowires are predicted to undergo a field-induced phase transition from a trivial to a topological superconductor, marked by the closure and re-opening of the excitation gap, followed by the emergence of Majorana bound states at the nanowire ends~\cite{lutchyn_majorana_2010,oreg_helical_2010}.
Many local density-of-states measurements have reported signatures of the topological phase~\cite{lutchyn_majorana_2018}, however this interpretation has been challenged by alternative explanations~ \cite{kells_near-zero-energy_2012,prada_transport_2012,chun-xiao_andreev_2017,vuik_reproducing_2018,reeg_zero-energy_2018,penaranda_non-hermitian_2019,prada_from_2020}. 
Here, by measuring nonlocal conductance, we identify the closure of the excitation gap in the bulk of the semiconductor before the emergence of zero-bias peaks.
This observation is inconsistent with scenarios where zero-bias peaks occur due to end-states with a trivially gapped bulk, which have been extensively considered in the theoretical \cite{kells_near-zero-energy_2012,prada_transport_2012,chun-xiao_andreev_2017,vuik_reproducing_2018,reeg_zero-energy_2018,penaranda_non-hermitian_2019,prada_from_2020} and experimental \cite{lee_zero-bias_2012,lee_spin-resolved_2013,yu_non-majorana_2020} literature.
We observe that after the gap closes, nonlocal signals fluctuate strongly and persist irrespective of the presence of local-conductance zero-bias peaks. 
Thus, our observations are also incompatible with a simple picture of clean topological superconductivity~\cite{lutchyn_majorana_2010,oreg_helical_2010}.
This work presents a new experimental approach for probing the spatial extent of states in Majorana wires, and reveals the presence of a regime with a continuum of spatially extended states and uncorrelated zero-bias peaks.}

\begin{figure}[h]
    \centering
    \includegraphics[scale=0.8]{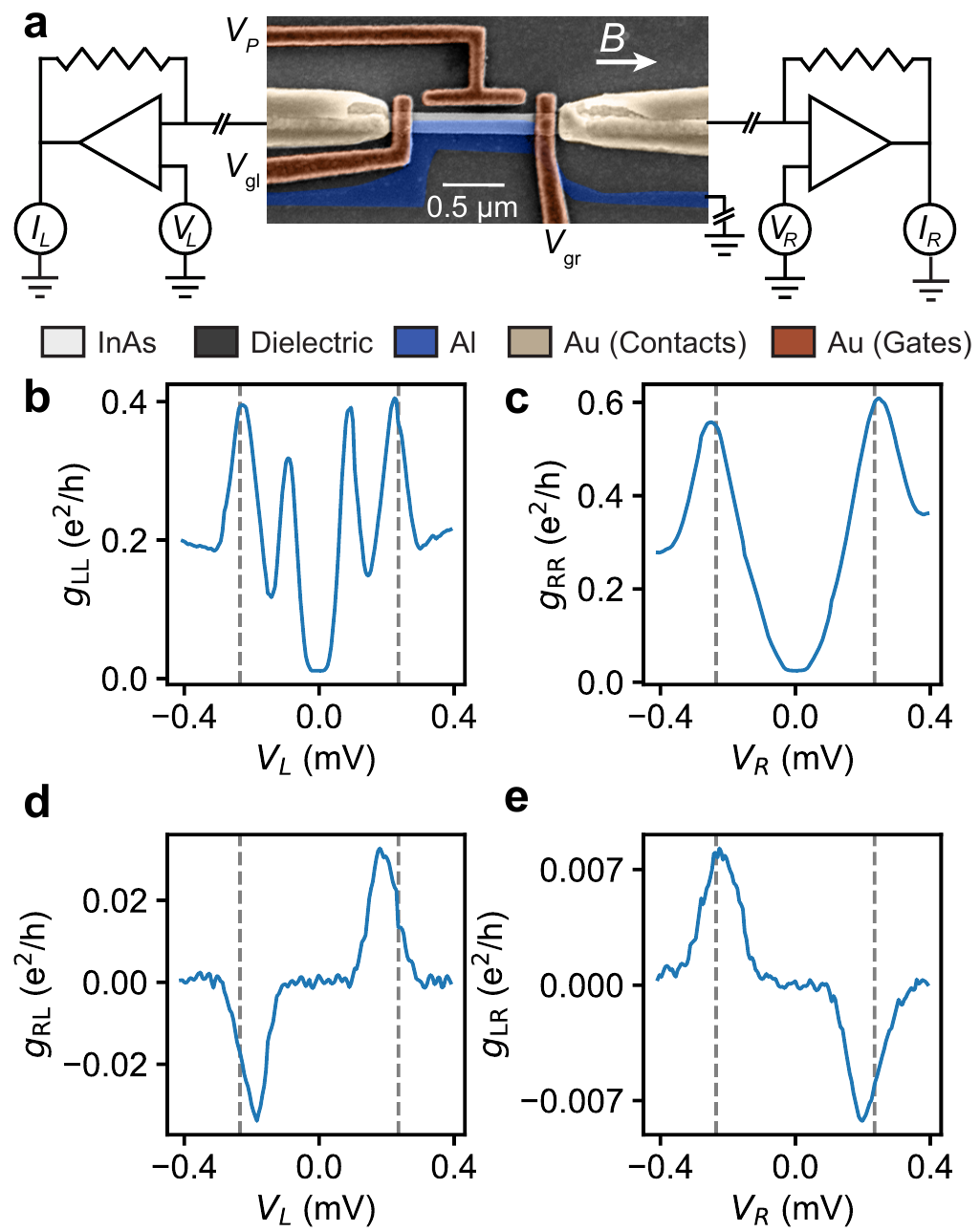}
    \caption{ \textbf{ Device layout and measurement setup. a,} Schematic of the measurement setup (outer) with a false-color scanning electron microscope image of the device (inner). Plunger gate $V_p$, tunnel barrier gates $V_{gl}$, $V_{gr}$, bias voltages $V_L$, $V_R$, and measured currents, $I_L$, $I_R$ are labeled. Line interruptions indicate the in-line filtering of the measurement setup, see methods. Direction of applied magnetic field, $B$, is indicated by the white arrow.
    Bias spectroscopy at $B=0$ for the full conductance matrix: \textbf{b,} $g_{ \text{LL}}$, \textbf{c,} $g_{ \text{RR}}$, \textbf{d,} $g_{ \text{RL}}$, and \textbf{e,} $g_{ \text{LR}}$. Dashed lines are a guide to the eye at $\pm$ 0.235 mV.}
    \label{fig:fig1}
\end{figure}

Topological superconductors are characterized by an excitation gap in the bulk~\cite{qi2011topological}, which only closes upon leaving the topological phase. In the ongoing experimental effort to control Majorana bound states in superconductor-semiconductor heterostructures, the evolution of the excitation gap is not well understood. Local density-of-states measurements by tunnelling spectroscopy have given varying evidence on the closure and re-opening of the excitation gap~\cite{mourik_signatures_2012,das_zero-bias_2012,deng_majorana_2016,vaitiekenas_effective_2018,nichele_scaling_2017}. 
Possible explanations for the absence of decisive signatures include weak coupling of bulk states to end-probes \cite{stanescu_to_2012,rainis_towards_2013}, finite-size effects \cite{mishmash_approaching_2016}, contributions from intrinsic Andreev bound states \cite{cole_proximity_2016} or contributions from end-localized Andreev bound states \cite{prada_transport_2012,chun-xiao_andreev_2017}.
The challenge of identifying the topological phase transition with local conductance measurements has led to proposals to detect bulk signatures of the topological transition using spin measurements~\cite{szumniak_spin_2017, serina_boundary_2018}, thermal conductance, nonlocal shot-noise \cite{akhmerov_quantized_2011}, Coulomb blockade spectroscopy~\cite{albrecht_exponential_2016,hansen_probing_2018}, quantum dot hybridization~\cite{prada_measuring_2017,clarke_2017} and, of particular relevance to this work, nonlocal Andreev rectification~\cite{rosdahl_andreev_2018}. 

Here, we explore the use of nonlocal conductance as a probe of bulk properties, finding experimental evidence for a field-driven transition into a regime with spatially extended states .
We investigate three-terminal superconductor-semiconductor hybrid devices shown in Fig.~\ref{fig:fig1}a. 
InAs nanowires are grown by molecular beam epitaxy with the selective area growth (SAG) method~\cite{vaitiekenas_selective-area-grown_2018,krizek_field_2018,lee_selective-area_2018}. A $9~\mathrm{nm}$ layer of Al is epitaxially deposited directionally on only one side of the nanowire \textit{in situ} to promote a transparent interface between the superconductor and the semiconductor~\cite{krogstrup_epitaxy_2015,chang_hard_2015}. 
Following growth, Al is selectively removed from the ends of the wire and the substrate, forming a superconducting terminal which extends from the nanowire to the bond pad. 
A Ti/Au ohmic contact is deposited at each end of the wire. Following a global HfOx dielectric deposition, electrostatic Ti/Au gates are deposited. The hybrid segment is $\sim$~1 \si\micro m long for the data presented in the main text; data for a $\sim$~2 \si\micro m long device, yielding similar results, is presented in the supplement.

As indicated in Fig.~\ref{fig:fig1}a, the superconducting Al lead is grounded during all measurements. 
Electrostatic gates $V_{\mathrm{gl}}$ and $V_{\mathrm{gr}}$ are adjusted to form tunnel barriers between the nanowire and the normal-conducting leads. 
A plunger gate, $V_p$, is used to tune the chemical potential in the nanowire.
The left (right) normal lead is biased with voltage $V_L$ ($V_R$) while measuring both currents $I_L$ and $I_R$~\cite{menard2019conductance}. Lock-in amplifiers are used to measure local differential conductance on the left $g_{\text{LL}}=dI_L/dV_L$ and right $g_{\text{RR}}=dI_R/dV_R$, as well as the differential nonlocal conductance $g_{\text{LR}}=dI_L/dV_R$ and $g_{\text{RL}}=dI_R/dV_L$.
The setup allows the $2\times2$ conductance-matrix, 
$$
g 
= 
\begin{bmatrix} 
g_{\text{LL}} & g_{\text{LR}} \\
g_{\text{RL}} & g_{\text{RR}}
\end{bmatrix},
$$
to be measured.

An example of the bias dependence of $g$ at zero magnetic field is shown in Fig.~\ref{fig:fig1}~b-e. 
In this instance, local conductance measurements on the left and right sides are qualitatively different [Fig.~\ref{fig:fig1}b,c].
Both local measurements have suppressed conductance at low bias and exhibit a pair of peaks near $V_{\mathrm{L,R}}\sim \pm 0.24~\mathrm{mV}$ (dashed vertical line), consistent with coherence peaks originating from the Bardeen-Cooper-Schrieffer (BCS) density of states in the hybrid wire. 
However, only $g_\mathrm{LL}$ exhibits an additional pair of peaks at lower bias voltage.
The additional peaks in $g_\mathrm{LL}$ indicate an Andreev bound state on the left side of the device, which is common in superconductor-semiconductor nanowires \cite{mourik_signatures_2012,deng_majorana_2016}, and can generate false-positive Majorana signatures~\cite{kells_near-zero-energy_2012,prada_transport_2012,chun-xiao_andreev_2017,vuik_reproducing_2018,reeg_zero-energy_2018,penaranda_non-hermitian_2019,prada_from_2020}.

In contrast to the local conductance, the two nonlocal conductances are qualitatively similar to each other~[Fig.~\ref{fig:fig1}d,e].
Both nonlocal conductances are negligibly small at low bias, consistent with the presence of an induced gap in the nanowire.
As bias is increased from zero, nonlocal conductance dramatically increases once a threshold bias is reached, characteristic of the presence of spatially extended modes above an induced gap.
The nonlocal conductance eventually reaches a peak, which roughly aligns with peaks in the local conductance, confirming the identification of an induced gap $\Delta_\mathrm{ind}\sim 0.24~\mathrm{meV}$.
At an even higher bias, the nonlocal conductance becomes negligibly small as expected due to the absorption of quasiparticles in the superconducting lead above the bulk Al gap, $\Delta_\mathrm{Al}$ \cite{rosdahl_andreev_2018}.
The disappearance of nonlocal conductance at $\Delta_\mathrm{Al}$ occurs close to the peak in nonlocal conductance at $\Delta_\mathrm{ind}$, suggesting $\Delta_\mathrm{ind} \sim \Delta_\mathrm{Al}$, as expected for a transparent, epitaxially matched Al/InAs interface~\cite{krogstrup_epitaxy_2015,chang_hard_2015,mikkelsen_hybridization_2018,antipov_effects_2018}.

Unlike the local measurement, the nonlocal signal does not show signatures of the sub-gap Andreev bound state. This is consistent with the expectation that nonlocal conductance is insensitive to the complicating effects of localized Andreev bound states for devices longer than the induced superconducting coherence length~\cite{rosdahl_andreev_2018}.
Motivated by the observed insensitivity to localized states, we now explore nonlocal conductance near the putative topological phase transition, where the role of localized bound states before the closure of the gap is an unsettled topic. 

\begin{figure}
    \centering
    \includegraphics[scale=0.41]{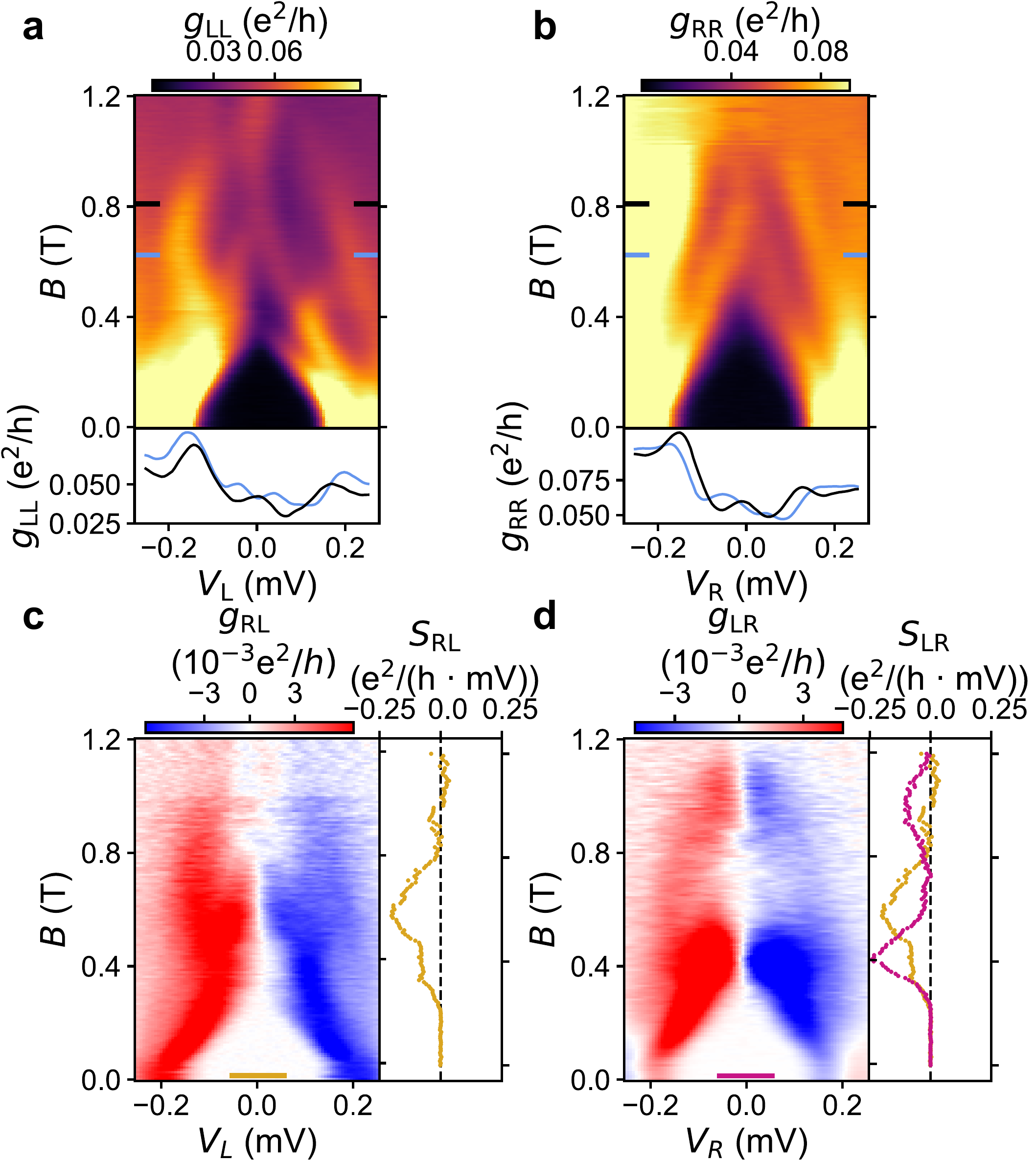}
    \caption{ \textbf{Magnetic-field dependence of the full conductance matrix.} 
    Local conductances \textbf{a}, $g_{\text{LL}}$ and \textbf{b}, $g_{\text{RR}}$ measured as a function of bias voltage and magnetic field $B$, with linecuts at $B=0.62~\mathrm{T}$ and $B=0.81~\mathrm{T}$ shown in the bottom panels.
    Nonlocal conductances \textbf{c}, $g_{\text{RL}}$ and \textbf{d}, $g_{\text{LR}}$ with nonlocal slope $S$ shown in right panel.
    $S$, was numerically computed over the bias window illustrated by the colored bar in \textbf{c} and \textbf{d} (Methods).
    For comparison, $S_\mathrm{LR}$ from \textbf{c} (orange) is overlaid in \textbf{d}.
    Data are taken at $V_p=-1.35~\mathrm{V}$. } 
    \label{fig:fig2}
\end{figure}

In Fig.~\ref{fig:fig2} the full conductance-matrix data are presented as a function of magnetic field $B$ applied parallel to the nanowire. The tunnel barrier and plunger gate voltages were tuned at finite $B$ to exhibit zero-bias peaks in both local conductances. Subsequently, the tunnel barrier gate voltages were held constant and the magnetic field and the plunger gate were swept. Both local conductances are suppressed at low bias at $B=$~0~T [main panels in Fig.~\ref{fig:fig2}a,b], and the size of the gap decreases with increasing magnetic field, eventually vanishing for $B \sim 0.4~\mathrm{T}$.
At a higher magnetic field, zero-bias peaks in the local conductance emerge on both sides of the device, which has never been reported in long superconductor/semiconductor nanowires to the best of our knowledge [bottom panels of Fig.~\ref{fig:fig2}a,b].
Zero-bias peaks in the local conductance at both ends of a device are a long-sought signature of Majorana bound states, which motivates further scrutiny of these features by nonlocal measurements.

In the nonlocal conductance, the gapped region around zero bias narrows as $B$ is increased [Fig.~\ref{fig:fig2}c,d].
For sufficiently large magnetic fields ($B\gtrsim 0.4$~T), the nonlocal conductance becomes non-zero around zero applied bias, which indicates a closure of the induced gap before the emergence of zero-bias peaks.
Our observation of the closure of the induced gap before the emergence of zero-bias peaks is inconsistent with a well-studied scenario involving quasi-Majorana modes that predicts the emergence of zero-bias peaks before the closure of the gap \cite{kells_near-zero-energy_2012,prada_transport_2012,chun-xiao_andreev_2017,vuik_reproducing_2018,reeg_zero-energy_2018,rosdahl_andreev_2018}.

Upon further increase of the magnetic field, nonlocal conductance is suppressed at low bias for $B\gtrsim 0.8~\mathrm{T}$ ($B\gtrsim 0.6~\mathrm{T}$) for the left (right) side. 
As previously mentioned, the nonlocal signal is also suppressed when the bias exceeds $\Delta_\mathrm{Al}$. The bias at which the signal is suppressed decreases with increasing $B$ and, at $B\gtrsim 1.2~\mathrm{T}$, the nonlocal signal mostly vanishes. This behavior is consistent with decreasing $\Delta_\mathrm{Al}(B)$, for higher magnetic fields.

To further quantify the behavior of the nonlocal conductance at low bias, we introduce the nonlocal slope, $S\equiv S_{\mathrm{ab}}$, given by $S_{\mathrm{ab}} = \frac{d^2 I_a}{d V_b^2}\Bigr\rvert_{V_b = 0}$, for $a=L,R$ and $b=R,L$.
$S$ is a metric for the strength of nonlocal conductance, which takes into account the predominantly anti-symmetric behavior of the nonlocal conductance in bias, giving $g_\mathrm{ab} \approx S_\mathrm{ab} V_b$ at low bias.

The right-side panels in Fig.~\ref{fig:fig2}c,d show $S_\mathrm{{RL}}$ and $S_{\mathrm{LR}}$, numerically extracted from the nonlocal conductances (see supplement).
For small magnetic fields, $S$ is zero, consistent with the previous qualitative observation that the nonlocal conductance is gapped at low field.
$S$ smoothly increases near $B\sim 0.35~\mathrm{T}$, initially exhibiting similar behavior on both sides of the device [Fig.~\ref{fig:fig2}d]. The appearance of gapless nonlocal conductance at a single characteristic field confirms the earlier interpretation of a closure in the induced gap in the entire hybrid wire.

Interestingly, $S$ changes non-monotonically with $B$.
$S_{\mathrm{RL}}$ is small over the interval $(0.8~\mathrm{T},1.2~\mathrm{T})$, whereas $S_{\mathrm{LR}}$ is small over the interval $(0.6~\mathrm{T},0.8~\mathrm{T})$. 
Thus, although the induced gap in the nanowire closes at a single characteristic field, $S$ fluctuates unevenly on both sides of the device at large fields and remains sizeable on at least one side when zero-bias peaks emerge [Fig.~\ref{fig:fig4}d right panel].

It is interesting to note that the nonlocal conductances,  $g_\mathrm{LR}$ and $g_\mathrm{RL}$, differ strongly at large magnetic fields.
In short devices it was previously found that $g_\mathrm{LR} \sim g_\mathrm{RL}$ even for large magnetic fields, indicating spatially homogeneous BCS charge~\cite{menard2019conductance}.
Following the same interpretation in the current device, the difference between $g_\mathrm{LR}$ and $g_\mathrm{RL}$ indicates a spatial inhomogeneity of the BCS charge, perhaps attributable to the longer device length or increased disorder~\cite{danon_nonlocal_2020}.

\begin{figure}
    \centering
    \includegraphics[scale=0.41]{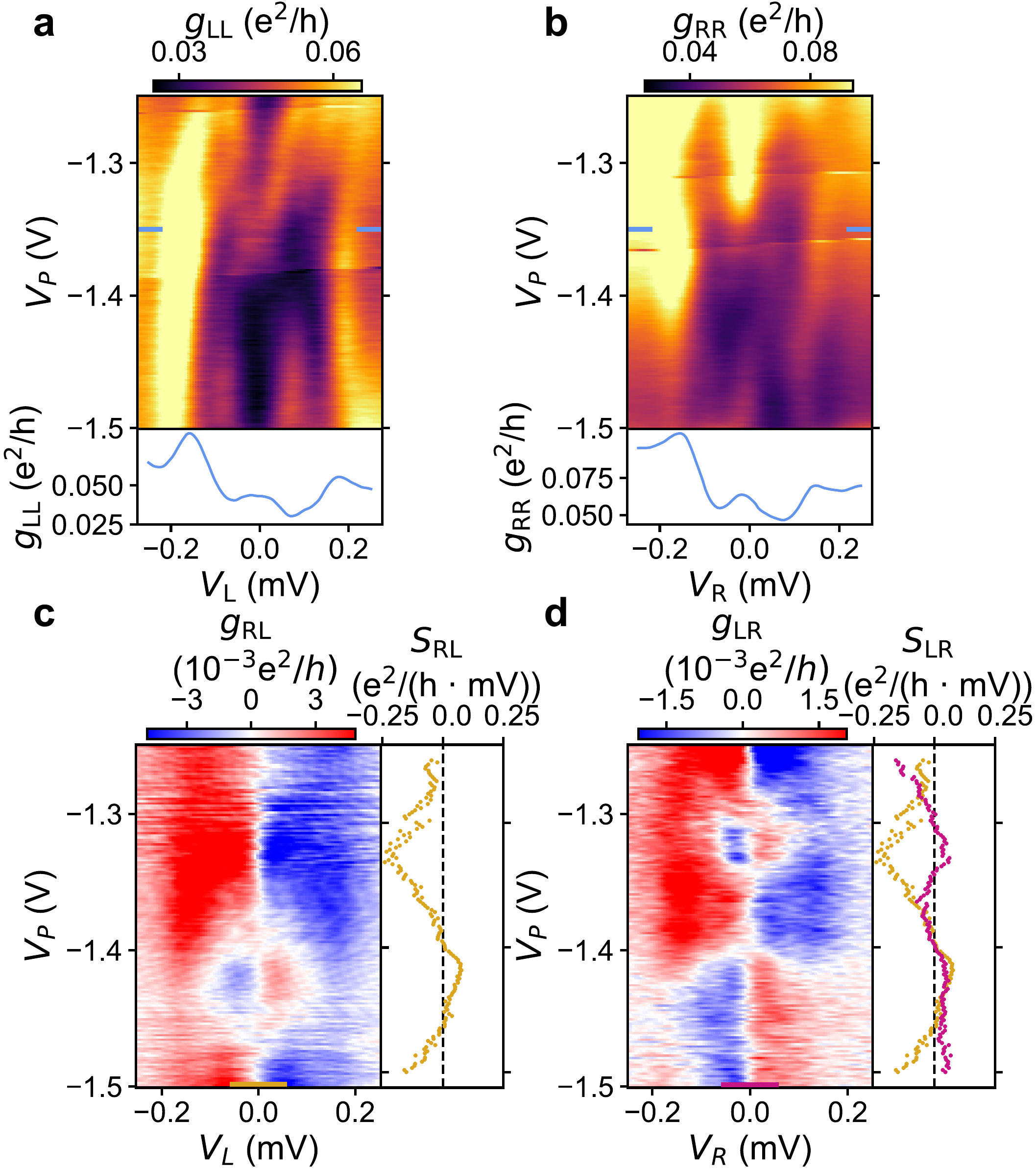}
    \caption{ \textbf{Gate-voltage dependence of the full conductance matrix at $B=0.7~\mathrm{T}$.} 
    Local conductances \textbf{a}, $g_{\text{LL}}$ and \textbf{b}, $g_{\text{RR}}$ measured as a function of bias voltage and plunger gate, with linecuts at $V_p=-1.35~\mathrm{V}$ shown in the bottom panels.
    Nonlocal conductances \textbf{c}, $g_{\text{RL}}$ and \textbf{d}, $g_{\text{LR}}$ with nonlocal slope $S$ shown in right panel.
    $S$ was numerically computed over the bias window illustrated by the colored bar in \textbf{c} and \textbf{d}.
    For comparison, $S_{\mathrm{RL}}$ from \textbf{c} (orange) is overlaid in \textbf{d}.    
    Data taken at $B=0.7~\mathrm{T}$.}
    \label{fig:fig3}
\end{figure}

The gate-voltage dependence of the zero-bias peaks and nonlocal conductance provides further information on the origin of these features.
Figure~\ref{fig:fig3} shows full conductance-matrix measurements as a function of plunger gate voltage. 
The scan was acquired by setting the magnetic field to 0.7~T, which is slightly lower than the value at which zero-bias peaks emerge in Fig.~\ref{fig:fig2}a,b [linecuts in the bottom panel in Fig.~\ref{fig:fig3}a,b], and then sweeping the plunger gate. 
In Fig.~\ref{fig:fig3}a, we see the state on the left side crosses through zero bias, which is not expected for a Majorana bound state.
The state on the right side of the wire, however, remains robust to voltage changes of over $100~\mathrm{mV}$~[Fig.~\ref{fig:fig3}b].
The overall dependence of the left and right zero-bias peaks on gate voltage are qualitatively different, which is not consistent with a pair of Majorana bound states in a clean topological superconductor.

Throughout the gate-voltage sweep, the nonlocal conductance fluctuates strongly, changing sign several times while remaining predominantly anti-symmetric in bias voltage [Fig. \ref{fig:fig3}c,d]. 
There is no apparent correlation between sign changes in the nonlocal conductance on either side.
Further, the low-bias nonlocal response persists even in the absence of zero-bias peaks.
Changing the plunger-gate by several volts does not remove the nonlocal conductance (see supplement), indicating that the presence of spatially extended modes is more robust than zero-bias  peaks.

Summarizing, the data in Fig.~\ref{fig:fig2} and Fig.~\ref{fig:fig3} are characterized by the generic emergence of fluctuating, low-bias nonlocal conductance above a characteristic field, and zero-bias peaks that are not end-to-end correlated.
The experimentally identified apparent gapless regime would be consistent with theoretical explanations where a gap exists but is immeasurably small, or where the characteristic length-scales for subgap states exceeds the device sizes we have studied ($1-2~\mathrm{\mu m}$).
These observations do not fit neatly within a known picture of a clean topological superconductor, since the uncorrelated behavior of zero-bias peaks and lack of a gap re-opening is inconsistent with a pair of Majorana bound states at the ends of the wire.
Gapless or small-gap regimes have been discussed in the context of orbital \cite{nijholt_orbital_2016,winkler2019unified} or disorder effects \cite{motrunich_griffiths_2001}.

\begin{figure}
    \centering
    \includegraphics[scale=0.41]{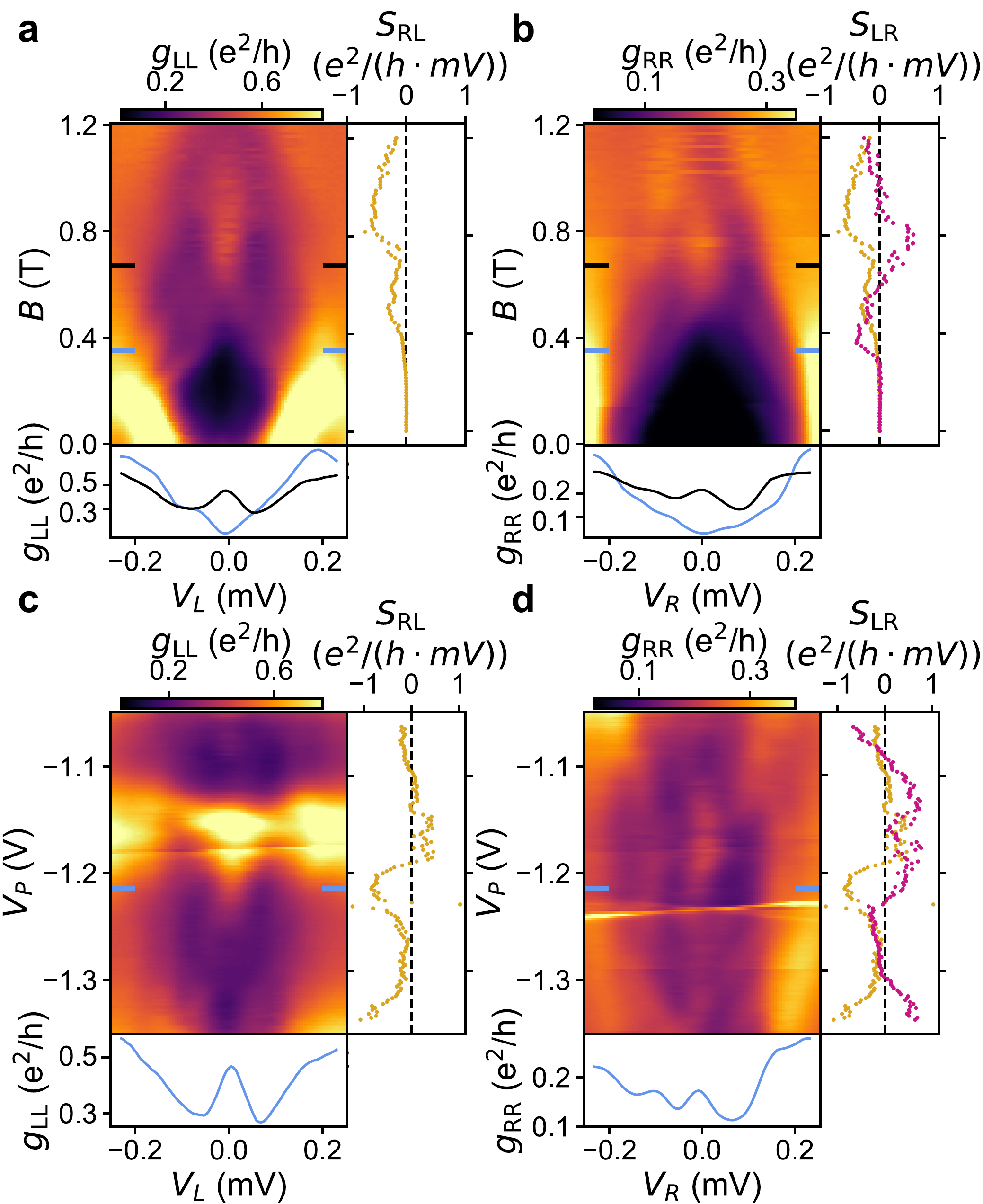}
    \caption{ \textbf{Full conductance matrix with correlated zero-bias peaks.}
    Local conductances \textbf{a,} $g_{\text{LL}}$ and \textbf{b}, $g_{\text{RR}}$ measured as a function of magnetic field and bias voltage with $V_p = -1.2~\mathrm{V}$. Linecuts taken at $B=0.32~\mathrm{T}$ and $B=0.64~\mathrm{T}$ and the bottom panels, and nonlocal conductance amplitude, $S$, in the right panels.
    Local conductances \textbf{c,} $g_{\text{LL}}$ and \textbf{d}, $g_{\text{RR}}$ measured as a function of gate voltage and bias voltage with $B=0.7~\mathrm{T}$. 
    Linecuts taken at $V_p=1.21~\mathrm{V}$ on the bottom panels, and nonlocal slope, $S$, on the right panels.
    For comparison, $S_{\mathrm{RL}}$ from \textbf{c} (orange) is overlaid in \textbf{d}.
    }
    \label{fig:fig4}
\end{figure}

Figure~\ref{fig:fig4} shows a different dataset, taken over a nearby plunger region but with different tunnel-barrier voltages. 
As magnetic field is increased, the gap in the local conductance is suppressed, and zero-bias peaks emerge on both sides of the device for an extended range in magnetic fields [Fig.~\ref{fig:fig4}a,b, linecuts in bottom panels].
The nonlocal slope, $S$, increases before the zero-bias peaks emerge, remaining sizable for large magnetic fields, qualitatively similar to the dataset in Fig.~\ref{fig:fig2}.  

Figure~\ref{fig:fig4}~c,d presents plunger-gate scans taken at $B=0.7~\mathrm{T}$, which is slightly after the emergence of zero-bias peaks in the local conductance. In both cases, low-bias peaks can be seen in line cuts [bottom panels in Fig.~\ref{fig:fig4}c,d], and remain near zero-bias for over $100~\mathrm{mV}$ plunger gate. 
A strong, bias-independent increase in conductance is seen in Fig.~\ref{fig:fig4}c at approximately $V_p=-1.15~\mathrm{V}$, which is compatible with resonant transmission due to coupling through a local state. 
The zero-bias peak does not split when it crosses this feature, suggesting that it is not coupled to the local state \cite{deng_majorana_2016,deng_nonlocality_2018}. 
The gate dependence of the zero-bias peaks on both sides is qualitatively similar, hinting that this particular zero-bias peak could be due to a single extended mode.
We emphasize that, in the absence of an excitation gap, spatially extended modes are not necessarily of a topological origin.
Further, we have established in previous work that in the presence of a dense, low-energy spectrum, statistically significant correlations can occur even in the presence of a topologically trivial excitation gap \cite{anselmetti_end-to-end_2019}.

Based on these observations, which have been reproduced in over 10 devices (see supplement for similar measurements in a $\sim$~2 \si\micro m long device), we conclude that the emergence of spatially extended states above a critical field, and a proliferation of zero-bias peaks, is a generic feature of our devices.
Further insight from theory is much needed to suggest potential origins of the this regime with zero-bias peaks.

We experimentally demonstrated that full conductance matrix measurements are capable of excluding two widely considered scenarios in hybrid one-dimensional systems: localized zero-energy states emerging before the gap closes and clean topological superconductivity. Localized zero-energy states emerging before the gap closes is inconsistent with our observation that apparently gapless nonlocal conductance generically precedes the appearance of peaks in local conductance. 
Clean topological superconductivity is excluded by the lack of end-to-end correlations and the absence of a gap reopening signature.
Since the lack of an excitation gap prohibits any progress on the path to topological quantum computation, we anticipate that nonlocal measurements will play an important role in benchmarking potential candidate systems for topological quantum computation.

\section{Methods}
The samples were measured in a dilution refrigerator with base temperature $T\sim$~50~mK. The resistance of the dilution refrigerator lines is $\sim 1.8\mathrm{k\Omega}$ and typical contact resistance for the SAG devices is $0.5-1\mathrm{k\Omega}$. Voltage divider effects have not been corrected for all the data presented in this manuscript. We note that the sample is operated in the tunneling regime such that the voltage drop at the tunnel barrier is always much larger than over the line resistance (fridge line resistance plus contact resistance).
Traces are acquired by sweeping $V_L$ with $V_R$ grounded, and then by sweeping $V_R$ with $V_L$ grounded. Voltage offsets are corrected for by both measuring DC currents and also making sure symmetric features like zero field coherence peaks appear at the same bias values. A constant voltage offset is assumed for two-dimensional scans. 
To extract $S$, numerical derivatives are computed from a polynomial fit is used over a bias window of $\pm$ 61 \si\micro~V centered around zero bias, see supplementary information.


\begin{acknowledgments}
\section{Acknowledgments}
We acknowledge insightful discussions with K.~Flensberg, E.~B.~Hansen, T.~Karzig, R.~Lutchyn, D.~Pikulin, E.~Prada, and R.~Aguado. This work was supported by Microsoft Project Q and the Danish National Research Foundation. C.M.M. acknowledges support from the Villum Foundation.
\end{acknowledgments}

\section{Author contributions}
R.K., S.G., G.C.G. and M.J.M. developed and grew the selective area growth samples.  A.P.H and L.C. designed the experiment. L.C. fabricated the devices and D.P. and G.C.M. performed the experiment with inputs from E.A.M., A.P., A.P.H. and L.C.. D.P., E.A.M., G.C.M., A.P., C.M.M., A.P.H., and  L.C. analyzed the data and prepared the manuscript. 

\section{Data availability statement}
The data that support the plots within this paper and other findings of this study are available from the corresponding author upon reasonable request.\newline

\section{Competing financial interest}
The authors declare no competing financial interests.\newline

\section{Additional information}
Reprints and permission information is available online at \url{www.nature.com/reprints}. Correspondence and requests for materials should be addressed to A.P.H. or L.C.

\begin{figure}
\includegraphics[page=1,trim={0.7in 0.7in 0.7in 0.7in},width=1.0\textwidth]{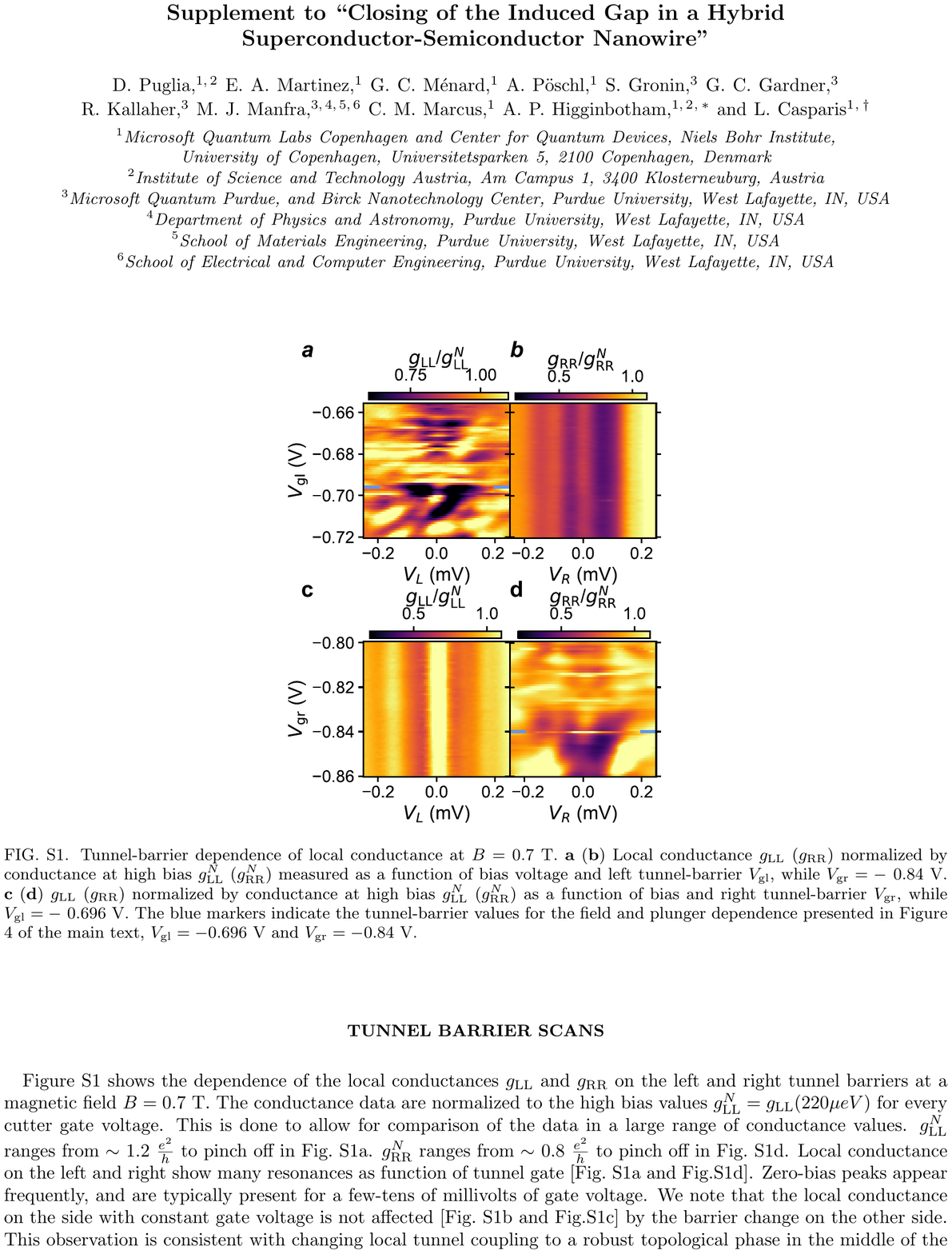}
\end{figure}
\begin{figure}
\includegraphics[page=2,trim={0.7in 0.7in 0.7in 0.7in},width=1.0\textwidth]{supplement}
\end{figure}
\begin{figure}
\includegraphics[page=3,trim={0.7in 0.7in 0.7in 0.7in},width=1.0\textwidth]{supplement}
\end{figure}
\begin{figure}
\includegraphics[page=4,trim={0.7in 0.7in 0.7in 0.7in},width=1.0\textwidth]{supplement}
\end{figure}
\begin{figure}
\includegraphics[page=5,trim={0.7in 0.7in 0.7in 0.7in},width=1.0\textwidth]{supplement}
\end{figure}

\end{document}